\documentclass[twocolumn,showpacs,floatfix,prl]{revtex4}
\usepackage{amssymb}
\usepackage{amsmath}
\usepackage{graphicx}
\usepackage{dcolumn}
\usepackage{bm}
\usepackage[usenames]{color}

\begin{document}

\title{Crossover from Super- to Sub-Diffusive Motion and Memory Effects in 
Crystalline Organic Semiconductors}
\author{G.~De~Filippis$^1$, V.~Cataudella$^1$, A.~S.~Mishchenko$^{2}$, 
N.~Nagaosa$^{2,3}$, A.~Fierro$^4$, and A.~de~Candia$^5$}
\affiliation{$^1$SPIN-CNR and Dip. di Fisica - Universit\`{a} di Napoli
Federico II - I-80126 Napoli, Italy \\
$^2$RIKEN Center for Emergent Matter Science (CEMS), Wako, Saitama 351-0198, Japan\\
$^3$Department of Applied Physics, The University of Tokyo,
7-3-1 Hongo, Bunkyo-ku, Tokyo 113-8656, Japan\\
$^4$SPIN-CNR Complesso Universitario di Monte S. Angelo- I-80126 Napoli, Italy \\
$^5$INFN, SPIN-CNR and Dip. di Fisica - Universit\`{a} di Napoli
Federico II - I-80126 Napoli, Italy 
}


\begin{abstract}

The transport properties at finite temperature 
of crystalline organic semiconductors are investigated, 
within the Su-Schrieffer-Heeger model, 
by combining exact diagonalization technique, 
Monte Carlo approaches, and 
maximum entropy method. The temperature-dependent mobility data measured in single crystals of 
rubrene are successfully reproduced: a 
crossover from super- to sub-diffusive motion occurs in the range  
$150 \leq T \leq 200$ K, where the mean free path becomes of the order of the lattice parameter 
and strong memory effects start to appear. 
We provide an effective model which can successfully explain low frequencies features of the 
absorption spectra. The observed response to slowly varying electric field 
is interpreted by means of a simple model where 
the interaction between the charge carrier and lattice polarization modes is simulated by 
a harmonic interaction between a fictitious particle and an electron embedded in a viscous fluid. 

\end{abstract}

\maketitle

Small molecule organic semiconductors, crystals of small molecules held together by van der 
Waals forces, 
are the focus of an intensive research activity being the material basis for 
the organic electronics, and in particular for the plastic electronics, a rapidly developing 
field\cite{refsem}. 
Because of the weak van der Waals intermolecular bonding, 
there is a small overlap between the electronic orbitals 
of these small molecules leading to narrow electronic bands: the transfer integral $t$ turns out to 
be about $100$ meV\cite{tro,corop}. 
At the same time 
the electron-phonon interaction (EPI) plays a crucial role\cite{han} and, now, it is well 
established that it stems from Peierls's coupling mechanism\cite{zoli}. 
EPI exhibits a strong momentum 
dependence, and, given the pronounced anisotropy of these compounds\cite{poz}, typically the coupling 
of the electrons with the lattice vibrations is described by using a one-dimensional tight-binding 
model involving Einstein phonons with the lattice displacements 
affecting the electronic hopping integral\cite{tro,tro1} (Su-Schrieffer-Heeger coupling\cite{su}).  
However the charge transport understanding in organic
semiconductors remains limited. Indeed from an experimental point of view ultrapure crystals of
pentacene or rubrene exhibit: 1) actived transport at low temperatures\cite{poz,poz1,mor} 
(up to $\simeq 160$ K); 2) 
a band-like mobility up to room temperature, i.e. the mobility
decreases as $T^{- \alpha}$ with $\alpha \simeq 2$\cite{poz1,tro2}. 
At the same time, at room temperature, 
optical absorption spectra are characterized by a broad peak centered around $40$ meV\cite{poz2}, 
reminiscent of disordered systems in the insulating phase. 
It has been shown that the
rapid drop of the mobility below $160$ K is due to the crossover to the trap dominated
regime (extrinsic disorder). Measurements of the transverse Hall conductivity\cite{poz1,tak} 
allowed to 
extract the intrinsic, trap-free mobility that increases always with cooling 
(shallow traps do not contribute to the Hall voltage since 
the Lorentz force is zero for these charge carriers).
It remains to explain the puzzle regarding
the simultaneous presence of the signature of intrinsic band-like transport (power law dependence
of mobility vs temperature) and localized states (absorption spectra feature broad peak
centered at finite frequency, whereas in the Drude model the optical conductivity (OC) 
shows a maximum 
at $\omega=0$). The starting point of several approaches present in literature is the mechanical 
softness of these compounds: the characteristic energy of the lattice 
transverse modes, $\omega_0$, 
is about $5$ meV that is much less than the electronic transfer integral $t$\cite{corop,tro3}. 
Then the idea is to use the adiabatic approximation: the phonon variables are treated classically 
and the electron contribution to the partition function is calculated at fixed lattice displacements 
neglecting the retardation effects\cite{catau}. 
In the one-dimensional case the problem turns out to be equivalent 
to that of a particle in presence of an off-diagonal disorder which provides 
a vanishing mobility (Anderson localization\cite{ander}). In order to overcome 
this difficulty the fluctuations of the lattice vibrations, completely neglected in the adiabatic 
approximation (it becomes exact when $\omega_0  \rightarrow 0$ and the ionic mass 
$M  \rightarrow \infty$ keeping $k=M \omega_0^2$ $constant$\cite{def}), 
have to be taken into account. 
To this aim it has been proposed: i) to employ mixed quantum-classical 
simulations based on the Ehrenfest coupled equations\cite{tro3}; ii) 
to neglect vertex corrections in the 
OC calculation\cite{ciu}; iii) to introduce an {\it ad hoc} energy broadening of the 
system energy levels\cite{catau}. 
All these approximated approaches, although able to reproduce a power law 
dependence of mobility vs temperature, do not restore the conductivity values quantitatively. 
It is clear that transport properties crucially depend on the exact dynamic of the lattice at 
long times. In this paper we go beyond the adiabatic approximation: 
both lattice and electronic degrees of freedom obey exactly (from a numeric point of view) 
quantum dynamics. We combine exact diagonalization technique\cite{giu}, 
diagrammatic\cite{andrey} and world-line\cite{antonio} Monte Carlo approaches, and
maximum entropy method\cite{max} to obtain an unbiased result. 
The model Hamiltonian is given by:
\begin{eqnarray}
H=\sum_{k} (\epsilon_{k}c_{k}^{\dagger}c_{k} + \omega_0b_{k}^{\dagger}b_{k})
+ \sum_{q,k} (M_{q,k}c_{k+q}^{\dagger} 
c_{k}b_{q}+h.c.),
\label{hamiltonian}
\end{eqnarray}
where $\epsilon_{k}=-2 t \cos(ka)$ denotes the electron band with 
hopping $t$, $c_{k}^{\dagger }$ ($b_{k}^{\dagger }$) represents the momentum electron (phonon)
creation operator and Einstein optical phonons have frequency $\omega_0$. EPI vertex is:  
$M_{q,k}=2 i \alpha t/\sqrt{N} (\sin(k+q)a-\sin(ka))$ describing the 
transfer integral modulation on the distance between nearest neighbours with strength $\alpha$ 
($N$ denotes the number of sites and $a$ is the lattice parameter).
We use units such that $\hslash=e=k_B=1$, where $e$ is the electronic charge and 
$k_B$ is the Boltzmann constant. 
We shall study the system by assuming values of the parameters typical of 
single crystal organic semiconductors, taking rubrene as a case study\cite{tro1}. 
We set: $a=7.2$ \AA, 
$\omega_0=0.05t$, $\alpha=0.092$ and $t=93$ meV.  
  
Within the linear response theory, the light absorption, at low densities, is proportional 
to the particle concentration. The proportionality constant is OC\cite{sha}:
\begin{eqnarray}
\sigma(z)=\frac{i}{z}\left(\Pi(z)- \Gamma \right),
\label{sigma}
\end{eqnarray}
where $z$ lies in the complex upper half-plane, $z=\omega+i\epsilon$ with
$\epsilon>0$, the quantity $\Gamma$ is:
\begin{eqnarray}
\Gamma= -\int_0^{\beta} ds  \left\langle j(s)j(0) \right\rangle,
\label{gam}
\end{eqnarray}
and $\Pi(z)$ represents the current-current
correlation function
\begin{eqnarray}
\Pi(z)=-i \int_0^{\infty} d\tau e^{i z \tau} \left\langle [j(\tau),j(0)] \right\rangle.
\label{pi}
\end{eqnarray}
In Eq.~(\ref{pi}) (Eq.~(\ref{gam})) $j(\tau)$ ($j(s)$) is the real-time (imaginary-time)
Heisenberg representation of the
current operator (see Supplemental Material), $[,]$ denotes the commutator, 
$\left\langle \right\rangle$ indicates the thermal average, and $\beta=1/T$.
The real part of OC 
($\operatorname{Re} \sigma(\omega)=\operatorname{Re} 
\sigma(\omega+i\epsilon)$, $\epsilon \rightarrow 0^{+}$) 
is related to the 
imaginary-time current-current correlation function: 
\begin{eqnarray}
\Pi(s)=\int_{-\infty}^{\infty} d\omega \frac {1}{\pi} \frac {\omega e^{-\omega s}}
{1-e^{\beta \omega}} \operatorname{Re} \sigma(\omega). 
\label{analytic}
\end{eqnarray}
The function $\Pi(s)$ has been calculated by using Diagrammatic\cite{andrey} 
and world-line\cite{antonio} Monte Carlo 
methods, checking that both approaches give the same results. The dynamical spectra, then, 
is extracted 
from the integral equation, Eq.~(\ref{analytic}), through the maximum entropy method\cite{max}. 
In particular
we used the Bryan's method choosing, as default model, the OC obtained through exact diagonalization
on a lattice of $20$ sites with periodic boundary conditions 
(at the investigated temperatures the mean free path (MFP) is
less than $6a$, so that such a small lattice provides a very good starting point). 

\begin{figure}
\flushleft
        \includegraphics[scale=0.38]{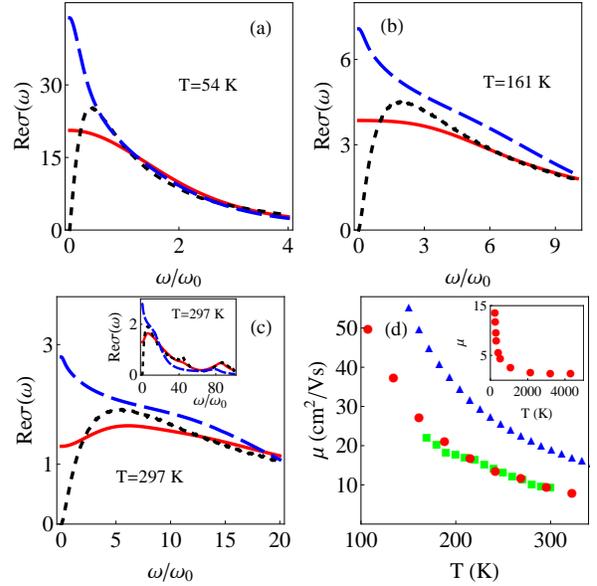}
        \caption{color online: (a), (b) and (c):
OC (in units of $a^2e^2/\hbar$) in different approximations: exact results (solid red line),
Boltzmann (long-dashed blue line) and adiabatic (short-dashed black line) approaches; (d):
temperature
dependence of the mobility in rubrene\cite{poz1} (green squares) 
compared with
exact results (red circles) and Boltzmann approach (blue triangles). In the inset
mobility (exact results in units cm$^2$/Vs) in a wider range of temperatures.}
\label{fig1}
\end{figure}

\begin{figure}
\flushleft
        \includegraphics[scale=0.38]{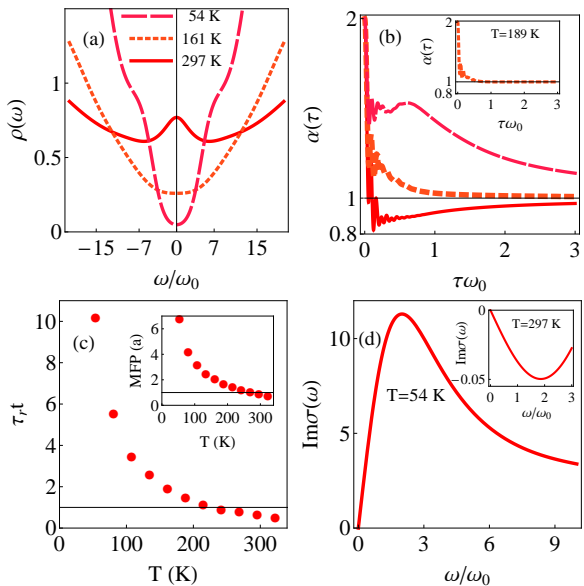}
        \caption{Resistivity (a) and diffusion exponent (b) at different temperatures (symbols
are the same in the two panels); (c): $\tau_r$ (in units of $1/t$) vs temperature (in the inset MFP
in units of $a$); (d): imaginary part of the conductivity at two temperatures.
}
\label{fig2}
\end{figure}

In Fig.1 we plot OC at three different temperatures, 
comparing exact results obtained from the numerical simulations 
with those of two limiting cases stemming from Boltzmann\cite{giu1} and 
adiabatic approaches\cite{catau}. 
It is evident that neither Boltzmann treatment, involving one-phonon scattering processes, nor 
the adiabatic approach, where electrons move in a static lattice and 
the mobility goes to zero at any temperature, are able to correctly describe the charge carrier 
dynamics at low frequencies\cite{note1}. Indeed only at very short times ($\omega \gg \omega_0$) 
OC behavior is well captured by the adiabatic treatment (see inset Fig.1c): 
at high 
frequencies the Franck-Condon principle can be invoked and the lattice is frozen during electron 
transitions between different energetic levels. On the other hand at long times ($\omega \leq 
\omega_0$) the lattice dynamic cannot be neglected and OC can be obtained only by treating 
correctly both electron and lattice vibration fluctuations. In Fig.1d the 
temperature-dependent mobility data measured in single crystals of
rubrene\cite{poz1} 
(intrinsic, trap-free mobility) are successfully compared with our numerical results: 
$\mu = \operatorname{Re} 
\sigma(\omega \to 0^+)/e$: our approach is able to recover both qualitatively and 
quantitatively the features of the 
absorption spectra at low frequencies. We emphasize (see inset of Fig.1d) the 
interesting mobility behavior at very high 
(but unphysical) temperatures indicating the saturation of the direct current conductivity 
at $T \simeq 1000$ K.

Also worthy of note is the remarkable behavior of the  
resistivity $\rho(\omega)=1/\sigma(\omega)$, 
as function of the frequency, shown in Fig.2a at different temperatures: 
at $T=54$ K there is a 
single absolute minimum at $\omega=0$; at $T=161$ K the curve is flat at this minimum; 
finally at $T=297$ K two minima develop at finite frequency. 
In order to investigate more deeply the microscopic physical 
mechanisms underlying this finding we have calculated the instantaneous diffusivity\cite{may}: 
\begin{eqnarray}
D(\tau)=\frac{1}{2} \frac {d \Delta x^2}{d \tau}= 
\int_{0}^{\infty} d\omega \frac {\operatorname{Re} \sigma(\omega) \sin(\omega\tau)}
{\pi \tanh(\frac{\beta\omega}{2})},
\label{diffusivity}
\end{eqnarray}
and the quantity $\Delta x^2= \left\langle (x(\tau)-x(0))^2 \right\rangle$, i.e. 
the mean-square displacement (MSD) 
of the position operator $x$. Locally, around $\tau_0$, MSD grows as $\tau^{\alpha(\tau_0)}$ with 
the diffusion exponent, $\alpha(\tau_0)$, equal to logarithmic slope of $\Delta x^2$: 
$\alpha(\tau_0)=\tau_0 D(\tau_0)/$ MSD$(\tau_0)$. 
In Fig.2b we plot the diffusion exponent vs the time $\tau$ (in units of $1/ \omega_0$) 
at different temperatures. At very short times 
$\alpha(\tau)$ is about $2$, i.e. the motion is ballistic independently on the temperature. 
However after a transient time the curves 
differ significantly. Indeed, at $T=\omega_0=54$ K the evolution is always super-diffusive 
($\alpha(\tau) > 1$) and approaches, only at long times, 
diffusive behavior, $\alpha(\tau) \rightarrow 1$. 
On the other hand, at $T=5.5 \omega_0=297$ K, there is a broad range of values of $\tau$ 
for which $\alpha(\tau) < 1$ signaling the onset of sub-diffusive motion. However, also here, 
at very large 
times, $\tau \gg 1/\omega_0$, the motion becomes diffusive. Finally in the range 
$150 \leq T \leq 200$ K (see also inset of Fig.2b) the ballistic motion is rapidly followed by 
the diffusion. Further insight into the problem 
is provided by the analysis of the optical relaxation time and 
the MFP. To this aim OC is written
in terms of the memory function $M(\omega)$\cite{mori}:
\begin{eqnarray}
\sigma(\omega)=-i \frac{\Gamma}{\omega+i M(\omega)},
\label{memory}
\end{eqnarray}
At $\omega=0$ the function $M$ is real and determines the reciprocal of the
optical relaxation time $1/\tau_r$, so that the mobility turns to be:
$\mu=- \tau_r \Gamma$. This last relation allows to extract $\tau_r$ (in SSH model the 
quantity $\Gamma$ is equal to the sum of 
thermal averages of the first and third term of the 
Hamiltonian Eq.~(\ref{hamiltonian})). 
The MFP is defined by: MFP $=v \tau_r$, and a rough estimate of the
average velocity $v$ can be obtained by $v \simeq \sqrt{\Pi(\tau=0)}$.
In Fig.2c (inset of Fig.2c) we plot the temperature dependence of $\tau_r$ (MFP). 
By increasing $T$, $\tau_r$ (MFP) decreases and 
becomes of the order of $1/t$ (the lattice parameter) just around $200$ K. The crossover 
from super- to sub-diffusive motion signals then the onset of the incoherent motion regime, where 
the charge carriers are strongly scattered in the real space and the band-like picture breaks down. 
Interestingly the imaginary part of the conductivity, at low frequencies, changes sign across 
the crossover (see Fig.2d) becoming negative when $T > 200$ K.   

It is possible to show that the subdiffusion is a direct consequence of the memory effects. 
To this aim we note that the Mori formalism allows to reformulate, in an exact way, 
the Heisenberg equation of motion of any dynamical variable in terms of a generalized Langevin 
equation\cite{mori1}. 
It is possible by introducing a Hilbert space of operators (whose 
invariant parts are set to be zero) where the inner product is 
defined by:
\begin{eqnarray}
(A,B)= \frac {1}{\beta} \int_{0}^{\beta} \left\langle e^{sH} A^{\dagger} e^{-sH} B \right\rangle ds.
\label{inner}
\end{eqnarray}
In particular the current operator obeys the equation:
\begin{eqnarray}
\frac{dj}{d\tau}=- \int_{0}^{\tau} M(\tau-r) j(r) dr + f(\tau),
\label{morij}
\end{eqnarray}
where the quantity $f(\tau)$ represents the ''random force'', that is, at any time, 
orthogonal to $j$ and 
is related to the memory function $M$ by the fluctuation-dissipation formula. The solution of this 
equation can be expressed as $j(\tau)=\Sigma(\tau)j+\tilde{j}(\tau)$, i.e. 
$\Sigma(\tau)=(j(\tau),j)/(j,j)$ describes the time 
evolution of the projection of $j(\tau)$ on the axis parallel to $j$ 
and represents the relaxation of the 
current operator (it is related to the Fourier transform of the real part of OC), 
whereas $\tilde{j}(\tau)$ is always orthogonal to $j$. In Fig.3a 
and Fig.3b we compare $\operatorname{Re} 
\Sigma(\tau)$ and the memory function $M(\tau)$ at $T=54$ K and 
$T=297$ K. The plots point out 
that while at low T the relaxation occurs on a timescale very much longer than the 
one characteristic 
of the memory function, at high T, where subdiffusion sets in, the two timescales are of 
the same order. It is a clear indication of presence of strong memory effects 
(breakdown of the Markovian approximation), that arise 
at $T >200$ K. 

\begin{figure}
\flushleft
        \includegraphics[scale=0.38]{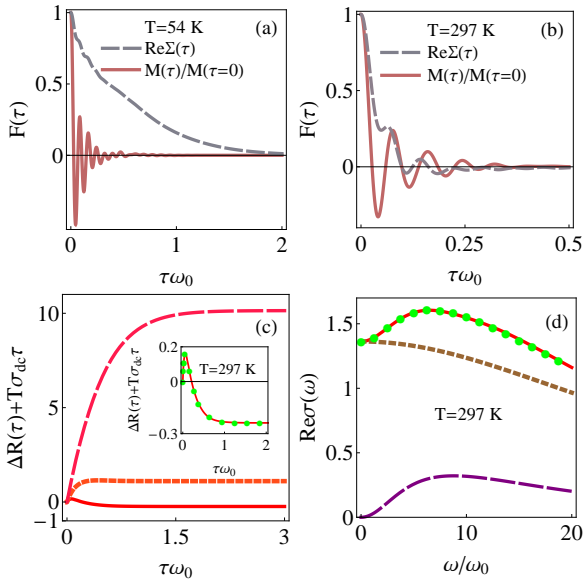}
        \caption{(a) and (b): comparison between current operator relaxation function 
and memory function at two temperatures
($F(\tau)$ stands for the dimensionless quantities $\operatorname{Re} \Sigma(\tau)$ and 
$M(\tau)/M(\tau=0)$); 
(c) relaxation function of the polarisation 
(symbols as in panel of Fig.2a) at different temperatures (in the inset comparison with the model 
(green circles));
(d): $\operatorname{Re} \sigma(\omega)$: exact results (red solid line); 
Drude-Lorentz contribution (DLC) (long-dashed purple line) (note that 
mobility of DLC is zero); 
Drude contribution (DC) (short-dashed brown line); DCL+DC (green circles).}
\label{fig3}
\end{figure}

Particularly interesting is the time response of the polarisation operator $P$\cite{pola}. 
It obeys an equations similar to Eq.~(\ref{morij}). 
We indicate with $R(\tau)$ the function $(P(\tau),P)$, that 
is related to the relaxation function of $P$, and with $\chi(\tau)$ the dielectric 
susceptibility (it is obtained by replacing $j(\tau)$ with $P(\tau)$ in Eq.~(\ref{pi})), that 
provides the response of $P$ to a weak external field. It is possible to show that 
$\frac{dR}{d\tau}=-\frac{1}{\beta} \chi(\tau)$ and $\chi(z)=i\sigma(z)/z$, so that the knowledge 
of $\sigma(z)$ fully determines the system response. In particular, if we assume a Drude model 
for the charge carriers, i.e. $\sigma(z)=\frac{\sigma_{dc}}{\tilde {\tau}_r} \frac{i}
{(z+i/\tilde {\tau}_r)}$ 
($\sigma_{dc}=\operatorname{Re} \sigma(\omega=0)$), 
$R(\tau)$ turns to be out: $\Delta R(\tau)= R(\tau)-R(\tau=0)=
-T \sigma_{dc} \left[ \tau+ \tilde {\tau}_r 
(e^{-\tau/\tilde {\tau}_r}-1) \right ]$. In Fig.3c 
we plot $G(\tau)=\Delta R(\tau)+T \sigma_{dc} \tau$ vs. $\tau$ at different temperatures. 
At $T=57$ K, $G(\tau)$ is very close to the function $T \sigma_{dc}\tau_r (1-e^{-\tau/\tau_r})$, 
i.e. Drude model with $\tilde{\tau}_r=\tau_r$ (best fit recovers just $\tilde{\tau}_r=\tau_r$) 
reproduces very well the behavior of $P$ relaxation at 
almost any time. By increasing temperature it is evident, 
from the plots in Fig.3b, that a Drude-like behavior is limited to a smaller range of $\tau$ values 
and a negative contribution appears, whose relaxation occurs on a longer time-scale. 
We found that a linear superposition of 
Drude- and Drude-Lorentz-like (damped harmonically bound particles (see Supplemental Material)) 
contributions represents a good fit of the function $G(\tau)$ 
at any temperature (see for example inset of Fig.3c). 
In any case it is worthy of note that
the best fit recovers $\tilde{\tau}_r=\tau_r$ at any temperature. 
While in the Drude model the memory 
effects are absent ($M(\tau)\propto \delta(\tau)$), 
in the Drude-Lorentz model the memory function 
is a superposition of two contributions: the first one is again $\propto \delta(\tau)$, whereas 
the other one is constant as function of the time $\tau$. 
As direct consequence of Eq.~(\ref{memory})
the mobility in Drude-Lorentz model is zero and the maximum of OC is located at finite frequency.
This explains the simultaneous presence of the signature of localized states (Drude-Lorentz
model) and intrinsic band-like transport (stemming from the Drude term).
Indeed Fig.3d shows 
that the linear superposition of two (Drude- and Drude-Lorentz-like) contributions (with 
parameters fixed by the best fit procedure for $G(\tau)$) 
is able to successfully describe the OC low frequencies behavior. 
Here the response can be described by simulating the
interaction between the charge carrier and lattice polarization modes by a harmonic interaction
between a fictitious particle and an electron embedded in a viscous fluid. The center of mass
(relative) motion turns out to be well represented by a Drude-like (Drude-Lorentz-like) model
(see Supplemental Material).
It is worthy of note that the dynamics
of the relative motion of this effective system is underdamped below $200$ K and becomes
overdamped above $200$ K. 
We found that the Drude-Lorenz-like contribution turns to be out negligible at $T<150$ K. 
On the other hand, above $150$ K, by increasing $T$ the weight of the
Drude-Lorentz-like contribution grows: this explains the change of sign of $\operatorname{Im} 
\sigma(\omega)$
at low frequencies observed in Fig.2d. Indeed $\operatorname{Im} \sigma(\omega)$ 
describes the current transport 
out of phase with the external field. $\operatorname{Im} \sigma(\omega)>0$ 
($\operatorname{Im} \sigma(\omega)<0$) 
in the Drude (Drude-Lorentz) model: at $T=200$ K there is a crossover 
from inductive behavior, due to the inertia of electrons, to capacitive behavior, due to the 
relative motion of electron-fictitious particle system. 
 
In conclusion, by combining exact diagonalization technique,
Monte Carlo approaches, and
maximum entropy method, we found that
a crossover from super- to sub-diffusive motion occurs in the range
$150 \leq T \leq 200$ K, where the relaxation time $\tau_r$ 
is of the order of $1/t$ ($\tau_r \ll 1/\omega_0$), 
the motion becomes incoherent, 
and strong memory effects start to appear. OC low frequencies features 
are well described by an effective model where the electron and 
a fictitious particle, embedded in a viscous fluid, harmonically interact with each other.

N.N. is supported by Grant-in-Aids for Scientific Research (S) (No. 24224009) 
from the Ministry of Education, Culture, Sports, Science and Technology (MEXT) 
of Japan and Strategic International Cooperative Program 
(Joint Research Type) from Japan Science and Technology Agency.


\begin{thebibliography}{99}
\bibitem{note1}
Boltzmann treatment provides results: i) quantitatively different (from the exact data)
by a factor of the order of two
and ii) exhibiting, in the most interesting range of temperatures, $200 < T < 300$ K, a power
law behavior, $\mu \simeq T^{-\alpha}$; however the index $\alpha$ turns out to be
$\simeq 1.5$, whereas both experimental and
theoretical data are characterized by $\alpha \simeq 2$.
\end{thebibliography}
\end{document}